# Unified Analysis of Transmit Antenna Selection/Space-Time Block Coding with Receive Selection and Combining over Nakagami-*m* Fading Channels in the Presence of Feedback Errors

Ahmet F. Coşkun and Oğuz Kucur

Gebze Institute of Technology, Electronics Engineering Department, Kocaeli, Turkey

{acoskun,okucur}@gyte.edu.tr

*Abstract*—Examining the effect of imperfect transmit antenna selection (TAS) caused by the feedback link errors on the performance of hybrid TAS/space-time block coding (STBC) with selection combining (SC) (i.e., joint transmit and receive antenna selection (TRAS)/STBC) and TAS/STBC (with receive maximal-ratio combining (MRC)-like combining structure) over Nakagami-*m* fading channels is the main objective of this paper. Under ideal channel estimation and delay-free feedback assumptions, statistical expressions and several performance metrics related to the post-processing signal-to-noise ratio (SNR) are derived for a unified system model concerning both joint TRAS/STBC and TAS/STBC schemes. Exact analytical expressions for outage probability and bit/symbol error rates (BER/SER) of binary and *M*-ary modulations are presented in order to provide an extensive examination on the capacity and error performance of the unified system that experiences feedback errors. Also, the asymptotic diversity order analysis, which shows that the diversity order of the investigated schemes is lower bounded by the diversity order provided by STBC transmission itself, is included in the paper. Moreover, all theoretical results are validated by performing Monte Carlo simulations.

I. INTRODUCTION

The devastating improvements that multi-antenna diversity schemes provided on the spectral efficiency [1] and reliability [2] of wireless communications systems over fading channels have caused these schemes to move on the focus of many researchers. Several work have been presented in the literature examining full-diversity reception and transmission techniques such as maximal-ratio combining (MRC) and space-time block coding (STBC) [2], [3] which function at the expense of the complexity caused by multiple radio-frequency (RF) chains. Subsequently, as expected, researches have been concentrated on the reduction of signal processing and hardware complexity in order to design more feasible and efficient wireless systems while maintaining the benefits of multi-antenna diversity. The most effective method to reduce the complexity of multi-antenna systems has been probably the antenna selection consisting of selection at the transmitter (transmit antenna selection (TAS)), receiver (receive antenna selection (RAS)) and both sides (joint transmit and receive antenna selection (TRAS)). The aim of these schemes is to maximize the post-processing signal-to-noise ratio (SNR) which brings along with full-diversity transmission and/or reception by performing the transmission and/or reception through a selected antenna subset. The antenna selection techniques can be employed at one side or both sides singly or in cooperation with other diversity techniques to form hybrid diversity schemes (HDSs).

The literature related to HDSs, which have been constructed by combining different diversity techniques at transmit and/or receive sides in order to obtain more diversity gain and/or reduce hardware complexity, have focused on TAS&MRC [4]-[6], TAS/STBC [7]-[9], TAS&generalized selection combining (GSC) [10] and joint TRAS [11], TAS/maximal-ratio transmission (MRT) [12], [13] and MRT&RAS [14] schemes. Also, TAS/STBC&selection combining (SC) (i.e., joint TRAS/STBC) scheme, which consists of combined TAS/STBC at transmit end and SC at receive end, has been considered in [15]-[18]. In [15], joint TRAS with space-time coding has been analyzed asymptotically for Rayleigh fading channels by using upper bounds and pairwise-error probability (PEP)-based approach. The error performance of a special case, that is based on performing the STBC transmission by selecting two transmit antennas (i.e., Alamouti Code (AC) [2]), has been examined in [16] for only binary phase-shift keying (BPSK) modulation and Rayleigh fading channels. Although authors of [17] have considered the same problem with arbitrary number of selected antennas, they have given results for quadrature phase-shift keying (QPSK) modulations and Rayleigh fading channels by using numerical integrations and not provided any closed-form expressions for the performance metrics. The outage probability performance of joint TRAS/STBC scheme has been examined for perfect feedback conditions in Nakagami-$m$ fading channels in [18]. The related literature does not include an extensive and exact performance analysis of joint TRAS/STBC scheme for several modulations, arbitrary system configurations and a generic fading model.

Also, in order to examine the feasibility of the diversity schemes and to see whether the investigated schemes are suitable for the real-world communications, researchers have also been dealing with the performance analysis of these schemes for practical issues such as channel estimation error (CEE), feedback delay (FD) and feedback error (FE). In [19], the effect of the Gaussian-modeled CEE on the outage probability performance of MRC structure has been investigated for Rayleigh fading channels. The performance of joint TRAS/STBC scheme in the presence of CEE has been analyzed in [20] for Rayleigh fading channels and some specific transmit and receive antenna configurations using approximations and PEP-based analysis method. The error performance of closed-loop transmit beamforming (TB) (i.e., MRT) scheme in Rayleigh fading channels has been analyzed in [21] in the presence of FD. Also, [14] presents the extensive performance analysis of hybrid MRT&RAS scheme in Nakagami-$m$ fading channels by considering both CEE and FD. The performances of closed-loop diversity schemes such as single TAS (i.e., selection diversity transmission (SDT)) and TB have been considered in [22] in the presence of FEs for Rayleigh fading channels by using upper bounds for BER of BPSK signals without providing any closed-form and exact BER expressions. Multi-input single-output (MISO) systems such as single TAS, TAS/AC (TAS/$G_2$-STBC), single TAS/power allocation (PA) and TAS/PA/AC have been analyzed considering FEs in [23] for Rayleigh fading channels and BPSK signals. The average BER expressions of TAS/AC scheme and TAS/PA/AC scheme with only four transmit antennas are derived for the selection of non-overlapping transmit antenna subsets which cannot guarantee the selection of the subset containing the best transmit antennas. The analyses provided in [22] and [23] deal with systems employing $n_T$ transmit antennas where $n_T = 2^k$, $k \in \mathbb{Z}^+$ that provides simplicity in mapping/de-mapping of feedback bits and the evaluation of a priori probabilities of each transmit antenna subset combination (TASC). In [24], authors have examined the downlink opportunistic scheduling scheme in Rayleigh fading channels in the presence of FE where the feedback quantization levels with arbitrary total numbers (not necessarily the power of 2) is considered. Also, the error performance analysis of joint TRAS/STBC scheme in the presence of FEs has been investigated in [33] for QPSK signals and Rayleigh fading case.

To the best of the authors' knowledge, there is no error performance analysis of joint TRAS/STBC scheme in independent identically distributed (i.i.d.) Nakagami-*m* fading channels for binary/*M*-ary modulations and even ideal feedback conditions. This paper, focusing on the unified analysis of hybrid joint TRAS/STBC scheme and TAS/STBC scheme (with MRC-like receiver) in the presence of FE in i.i.d. and flat Nakagami-*m* fading channels, makes the following specific contributions:

- We derive the probability density function (PDF) and the cumulative distribution function (CDF) of the output SNR.
- The system capacity is examined from the outage probability perspective.
- By using the conventional PDF-based and the moment generating function (MGF)-based analysis methods, the MGF of the output SNR and exact error probability expressions of binary and *M*-ary modulations such as BPSK, QPSK, differential BPSK (DBPSK), coherent and non-coherent binary frequency shift keying (CBFSK/NCBFSK), *M*-ary pulse and quadrature amplitude modulation (*M*-PAM/*M*-QAM) signals and the approximate error probability expressions of *M*-ary phase shift keying (*M*-PSK) signals (for $M \geq 8$) are derived for the erroneous feedback case in terms of a unified integral expression.
- The asymptotic diversity order analysis is carried out in order to derive the diversity order of the investigated HDSs.

- Also, the analytical performance results in the presence of the FE are validated with the help of Monte Carlo simulations.

Note that, the analysis related to joint TRAS/STBC scheme in erroneous feedback conditions is valid for the fading parameters of integer values ($m \in \mathbb{Z}^+$). Whereas, the analysis related to TAS/STBC scheme is valid for arbitrary values of $m \geq 1/2$ that satisfy the product of the fading parameter $m$ and the number of receive antennas ($n_R$) to be integer (i.e., $mn_R \in \mathbb{Z}^+$).

The remainder of this paper is organized as follows. In Section II, we present the system and channel model, briefly define the statistics of the output SNR by deriving the PDF and CDF expressions. In Section III, the outage probability analysis has been done in order to examine the system capacity. Section IV presents the derivation of the unified performance expression covering the MGF of the output SNR and exact (and approximate for *M*-PSK, $M \geq 8$) BER/SER performance expressions. The asymptotic diversity order analysis of the investigated hybrid systems is provided in Section V. In Section VI, we present some numerical results supported by simulations for outage probability and average error rates of binary and *M*-ary signals. Finally, Section VII draws conclusions about the analytical and numerical results.

## II. SYSTEM MODEL AND SNR STATISTICS

This paper focuses on two different HDSs with identical transmitter structure but different receiver structures which both have $n_T$ and $n_R$ antennas at the transmit and receive ends, respectively as depicted in Fig. 1. The transmitter side employs orthogonal STBC after determining the best available transmit antenna subset with $n_S$ ones out of $K = \binom{n_T}{n_S}$ combinations. At every STBC codeword transmission, the transmit antenna subset that maximizes the post-processing SNR at the receiver is activated while the other antennas are kept silent which results in reduced number of RF chains (i.e., reduced number of power amplifiers and complexity) at the transmitter. Whereas, the receiver side employs SC and MRC-like structure for joint TRAS/STBC and TAS/STBC cases, respectively.

The channel gains between the transmit antenna $j$ and the receive antenna $i$ are denoted by $h_{i,j}$ for $i = 1,2,\ldots,n_R$ and $j = 1,2,\ldots,n_T$ and the envelopes $|h_{i,j}|$ are assumed to be i.i.d. Nakagami random variates with fading parameter $m$ and squared mean $\Omega = E[|h_{i,j}|^2]$, where $E[\cdot]$ denotes the expectation operator. Before signal transmission, channel estimation and the selection of the transmit antenna subset (and the receive antenna for joint TRAS/STBC case) are performed by using pilot symbols. For joint TRAS/STBC, pilot symbols are sent by the transmit antennas, and the received signals of all receive antennas are monitored in turn via single RF circuit and only the receive antenna having the largest SNR is selected and the corresponding signal is fed to the single RF chain of the receiver for further processing. Whereas for TAS/STBC scheme, pilot symbols are monitored by all receive antennas and the optimal transmit antenna subset maximizing the post-processing SNR at the receive end is determined.

Under the assumptions that the channel estimation and the antenna selection processes are perfectly done, the antenna indices at both the transmit and receive sides are determined and the index of the TASC that consists of the best transmit antennas (unlike [23] in which the optimal transmit antenna subset is not guaranteed) is sent back to the transmitter through a low-rate feedback link which can be modeled as binary symmetric channel (BSC), as considered in [23]. The low-rate structure of the feedback link can be maintained by sending only the index of the TASC to the transmitter as the feedback information. This can be provided by defining a look-up table-like codebook (at both transmit and receive ends) consisting of the indices of all TASCs. The deteriorating structure of the BSC causes some errors (i.e., FE) at the delivery of the TASC index information (i.e., partial channel state information (CSI)) that results in degradation on the performance of closed-loop transmit diversity schemes [23]. In order to examine the effect of FE on the performance of the systems we consider in this paper, we carry out the analysis based on the BSC assumptions and the total error probability expression given as

$$P_s(p_e) = p_{CF} P_{s,CF} + p_{EF} P_{s,EF}(p_e). \tag{1}$$

In (1), $p_e$ denotes the average BER defined for each bit transmission over BSC. The feedback message should deliver $\eta = \lceil \log_2 K \rceil$ bits in order to represent $K$ different TASCs where the operator $\lceil \cdot \rceil$ denotes the smallest integer that is greater than the argument. But, since $\eta$ bits can represent $L = 2^\eta \geq K$ different states, FE may cause the proper TASC indices $(\mathbf{x}_1, \mathbf{x}_2, \ldots, \mathbf{x}_K)$ to be de-mapped to one of the improper TASC indices $(\mathbf{x}_{K+1}, \mathbf{x}_{K+2}, \ldots, \mathbf{x}_L)$ (which we call feedback failure (FF) in this paper) after the feedback transmission. Although many re-transmission or error control coding protocols (such as automatic repeat request (ARQ)) can be employed to overcome the FF, we consider that the transmitter arbitrarily selects and activates one of the proper TASCs in order not to reduce the transmission rate that would be brought along with the usage of a re-signaling structure. Thus, assuming that all proper TASCs are selected equally-likely in the presence of FF as in [24], we define the a priori probabilities for the correct feedback (CF) and erroneous feedback (EF) cases as

$$p_{CF} = \begin{cases} (1-p_e)^\eta + \dfrac{1}{K^2} \sum_{i=1}^{K} \sum_{j=0}^{L-K-1} p_e^{d_H(i,j)} (1-p_e)^{\eta - d_H(i,j)}, & L > K, \\ (1-p_e)^\eta, & L = K, \end{cases} \quad (2)$$

and $p_{EF} = 1 - p_{CF}$, respectively concerning also the case of $L > K$ unlike the references [22] and [23] which consider $K = 2^b, b \in \mathbb{Z}^+$. In (2), the term in the exponents $d_H(i,j) = d(\mathbf{x}_i, \mathbf{x}_{L-j})$ denotes the Hamming distance between the bit representations of the indices $\mathbf{x}_i$ and $\mathbf{x}_{L-j}$. In (1), $P_{s,CF}$ is the average SER expression related to the only correct TASC and

$$P_{s,EF}(p_e) = \sum_{\substack{\mathbf{c}_k \\ k=2,\ldots,K}} p_{\mathbf{c}_k} P_{s,EF}^{(\mathbf{c}_k)} = \frac{1}{K-1} \sum_{\substack{\mathbf{c}_k \\ k=2,\ldots,K}} P_{s,EF}^{(\mathbf{c}_k)} \quad (3)$$

is the average SER expression related to the EF case that is defined as the weighted sum of SERs $\left(P_{s,EF}^{(\mathbf{c}_k)}\right)$ related to $K-1$ wrong TASCs (denoted as $\mathbf{c}_k$, $k = 1, 2, \ldots, K$) with equal probability of occurrence (i.e., $p_{\mathbf{c}_k} = 1/(K-1)$). In (3), $\mathbf{c}_k = (n_1\ n_2\ \cdots\ n_{n_S})$, $n_{j+1} > n_j$, $j = 1, 2, \ldots, n_S - 1$, denotes the vector of the selected transmit antenna indices and each vector is assigned to a feedback symbol $\mathbf{x}_k$ such that the single

CF case is denoted by the first TASC ($\mathbf{c}_1 = (1\ 2\ \cdots\ n_S)$).

After the channel estimation, antenna selection and feedback processes, the STBC transmission results in an instantaneous SNR $\gamma_{i,j} = \frac{E_s}{n_S N_0 R_s}|h_{i,j}|^2$, $i = 1,2,\ldots,n_R$, $j = 1,2,\ldots,n_T$ and an average SNR $\bar{\gamma} = \frac{E_s \Omega}{n_S N_0 R_s}$ for each diversity branch where $R_s$, $E_s$ and $N_0$ denote the code rate of the utilized STBC, the average energy per symbol and the one sided power spectral density of additive white Gaussian noise (AWGN) at each receive antenna, respectively. The PDF and CDF expressions related to the instantaneous SNR of each branch for TRAS/STBC scheme have been obtained in [18] (also in [5], [6], [9], [11], [13] and [14]) as

$f(x) = \left(\frac{m}{\bar{\gamma}}\right)^m \frac{x^{m-1} e^{-xm/\bar{\gamma}}}{\Gamma(m)}, x \geq 0,$ and $F(x) = \frac{\psi(m, xm/\bar{\gamma})}{\Gamma(m)}, x \geq 0,$ respectively where $\Gamma(s) = \int_0^\infty t^{s-1} e^{-t} dt$, $Re(s) > 0$, denotes Gamma function [25, (8.310.1)] and $\psi(s,x) = \int_0^x t^{s-1} e^{-t} dt$, $Re(s) > 0$, denotes the incomplete Gamma function [25, (8.350.1)]. For TAS/STBC scheme, the PDF and CDF expressions of the transmit branch (through all receive antennas) have been given similarly in [9] as

$f(x) = \left(\frac{m}{\bar{\gamma}}\right)^{mn_R} \frac{x^{mn_R - 1} e^{-xm/\bar{\gamma}}}{\Gamma(mn_R)}, x \geq 0,$ and $F(x) = \frac{\psi(mn_R, xm/\bar{\gamma})}{\Gamma(mn_R)}, x \geq 0,$ respectively. Thus, we have unified the CDF and PDF expressions given above by the following expressions:

$$f_U(x) = \left(\frac{m}{\bar{\gamma}}\right)^{mg} \frac{x^{mg-1} e^{-xm/\bar{\gamma}}}{\Gamma(mg)}, x \geq 0, \tag{4}$$

$$F_U(x) = \frac{\psi(mg, x\, m/\bar{\gamma})}{\Gamma(mg)}, x \geq 0, \tag{5}$$

which can also be expressed as

$$F_U(x) = 1 - e^{-x\frac{m}{\bar{\gamma}}} \sum_{k=0}^{mg-1} \left(\frac{xm}{\bar{\gamma}}\right)^k \frac{1}{k!}, x \geq 0, \tag{6}$$

for integer values of the product $mg$, where $g = 1$ for joint TRAS/STBC and $g = n_R$ for TAS/STBC. The instantaneous output SNR for joint TRAS/STBC scheme will be the maximum of the receive branch SNRs:

$\gamma_{joint\ TRAS/STBC} = \max_{1 \leq i \leq n_R}\{\gamma_{joint\ TRAS/STBC}^{(i)}\}$. Thus, the CDF expression of the output SNR will be $F_{\gamma_{joint\ TRAS/STBC}}(x) = \prod_{i=1}^{n_R} F_{\gamma_{joint\ TRAS/STBC}^{(i)}}(x)$. Here, $F_{\gamma_{joint\ TRAS/STBC}^{(i)}}(x)$ is the CDF expression related to the instantaneous branch SNR $\gamma_{joint\ TRAS/STBC}^{(i)} = \sum_{k=1}^{n_S} Z_{n_k}^{(i)}$ where $Z_j^{(i)}, j = 1,2,\ldots,n_T$, denotes the sorted version of the branch SNRs such that the inequality $Z_1^{(i)} \geq Z_2^{(i)} \geq \cdots \geq Z_{n_T}^{(i)}$ holds true. Also, for TAS/STBC scheme, the output SNR will be $\gamma_{TAS/STBC} = \sum_{k=1}^{n_S} Y_{n_k}$ where $Y_j, j = 1,2,\ldots,n_T$, denotes the sorted version of the branch SNRs (after the MRC-like combining structure of STBC) such that the inequality $Y_1 \geq Y_2 \geq \cdots \geq Y_{n_T}$ holds true. Thus, the output SNR of both joint TRAS/STBC and TAS/STBC schemes can be unified by the SNR expression given as

$$\gamma_U = \max_{1 \leq v \leq N}\{\gamma^{(v)}\}, \qquad (7)$$

where the number of branches related to the receive selection is $N = n_R$ for joint TRAS/STBC and $N = 1$ (no selection) for TAS/STBC. The PDF of the branch SNR $\gamma^{(v)}$ given in (7) can be obtained by using the joint PDF given in [26, (2.2.3)] as

$$f_{(n_1)(n_2)\cdots(n_{n_S})}^{(v)}(x_1, x_2, \ldots, x_{n_S}) = n_T! \left\{\prod_{k=1}^{n_S} f_U(x_k)\right\} \prod_{k=1}^{n_S+1} \left\{\frac{[F_U(x_{k-1}) - F_U(x_k)]^{n_k - n_{k-1} - 1}}{(n_k - n_{k-1} - 1)!}\right\}, \qquad (8)$$

where $x_k \geq x_{k+1}$, $x_0 = +\infty$, $x_{n_S+1} = 0$, $n_0 = 0$ and $n_{n_S+1} = n_T + 1$. By using binomial expansion over the difference of CDF expressions given in terms of $F_U(x)$ as

$$[F_U(x_{k-1}) - F_U(x_k)]^{n_k - n_{k-1} - 1} = \sum_{p_k^{(v)}=0}^{n_k - n_{k-1} - 1} c_{p,k}^{(v)} [F_U(x_{k-1})]^{p_k^{(v)}} [F_U(x_k)]^{n_k - n_{k-1} - 1 - p_k^{(v)}} \qquad (9)$$

where $c_{p,k}^{(v)} = (-1)^{n_k - n_{k-1} - 1 - p_k^{(v)}} \binom{n_k - n_{k-1} - 1}{p_k^{(v)}}$ and also by substituting (4) and the binomial expanded expression of $[F_U(x_k)]^{d_k^{(v)}}$

$$[F_U(x_k)]^{d_k^{(v)}} = \left(1 - e^{-x_k m/\bar{\gamma}} \sum_{r=0}^{mg-1} (x_k m/\bar{\gamma})^r \frac{1}{r!}\right)^{d_k^{(v)}} = \sum_{t_k^{(v)}=0}^{d_k^{(v)}} \sum_{r_k^{(v)}=0}^{t_k^{(v)}(mg-1)} c_{tr,k}^{(v)} x_k^{r_k^{(v)}} e^{-t_k^{(v)} x_k m/\bar{\gamma}} \quad (10)$$

into (9), the joint PDF in (8) can be expressed as

$$f_{(n_1)(n_2)\cdots(n_{n_S})}^{(v)}(x_1, x_2, \ldots, x_{n_S}) = \frac{c_0 (m/\bar{\gamma})^{mg n_S}}{\Gamma(mg)^{n_S}} \sum_{P^{(v)}} \sum_{T^{(v)}} \sum_{R^{(v)}} \prod_{k=1}^{n_S} \left\{ c_{p,k}^{(v)} c_{tr,k}^{(v)} x_k^{mg+r_k^{(v)}-1} e^{-x_k m(1+t_k^{(v)})/\bar{\gamma}} \right\}. \quad (11)$$

In (10) and (11), the parameters are defined as $c_0 = n_T!/\prod_{k=1}^{n_S+1}\{(n_k - n_{k-1} - 1)!\}$, $d_k^{(v)} = n_k - n_{k-1} - 1 - p_k^{(v)} + p_{k+1}^{(v)}$ (note that $p_{n_S+1}^{(v)} = 0$) and $c_{tr,k}^{(v)} = \binom{d_k^{(v)}}{t_k^{(v)}} (-1)^{t_k^{(v)}} \beta_{r_k^{(v)}}(t_k^{(v)}, mg)$ where $\beta_{r_k^{(v)}}(t_k^{(v)}, mg)$ denotes the multinomial coefficients [25, (0.314)], [11]. The summations defined with the indices $\boldsymbol{P}^{(v)}$, $\boldsymbol{T}^{(v)}$ and $\boldsymbol{R}^{(v)}$ in (11) all denote $n_S$-fold summations with the indices $p_k^{(v)}$, $t_k^{(v)}$ and $r_k^{(v)}$, respectively. Since the multivariate structure of the PDF expression in (11) is analytically challenging, the Laplace transform and the inverse Laplace transform pair is applied over (11) in order to provide a more simplified representation. The Laplace transform of (11) can be obtained by substituting (11) into the multivariate Laplace integral as

$$H^{(v)}(s) = E\left[e^{-s(x_1+x_2+\cdots+x_{n_S})}\right] = \frac{c_0 (m/\bar{\gamma})^{mg n_S}}{\Gamma(mg)^{n_S}} \sum_{P^{(v)}} \sum_{T^{(v)}} \sum_{R^{(v)}} \prod_{k=1}^{n_S} \left\{ c_{p,k}^{(v)} c_{tr,k}^{(v)} \right\} \int_{x_{n_S}=0}^{\infty} x_{n_S}^{\mu_{n_S}^{(v)}-1} e^{-x_{n_S} b_{n_S}^{(v)}}$$

$$\times \int_{x_{n_S-1}=x_{n_S}}^{\infty} x_{n_S-1}^{\mu_{n_S-1}^{(v)}-1} e^{-x_{n_S-1} b_{n_S-1}^{(v)}} \cdots \int_{x_1=x_2}^{\infty} x_1^{\mu_1^{(v)}-1} e^{-x_1 b_1^{(v)}} dx_1 dx_2 \cdots dx_{n_S}, \quad (12)$$

where $\mu_k^{(v)} = mg + r_k^{(v)}$ and $b_k^{(v)} = \left(s + \frac{m(1+t_k^{(v)})}{\bar{\gamma}}\right)$. The integrals could be performed sequentially in order to obtain the closed-form expression of (12). But it is possible to derive the Laplacian expression by manipulating [27, (13)] as

$$H^{(v)}(s) = \frac{c_0(m/\bar{\gamma})^{mgn_S}}{\Gamma(mg)^{n_S}} \sum_{\boldsymbol{P}^{(v)}} \sum_{\boldsymbol{T}^{(v)}} \sum_{\boldsymbol{R}^{(v)}} \prod_{k=1}^{n_S}\left\{c_{p,k}^{(v)} c_{tr,k}^{(v)}\right\} \sum_{l_1^{(v)}=0}^{\mu_1^{(v)}-1} \frac{\Gamma(\mu_1^{(v)})}{l_1^{(v)}!} \frac{\left(s+a_1^{(v)}\right)^{-u_1^{(v)}}}{n_S^{-u_1^{(v)}}}$$

$$\times \sum_{l_2^{(v)}=0}^{l_1^{(v)}+\mu_2^{(v)}-1} \frac{\Gamma(l_1^{(v)}+\mu_2^{(v)})}{l_2^{(v)}!} \frac{\left(s+a_2^{(v)}\right)^{-u_2^{(v)}}}{(n_S/2)^{-u_2^{(v)}}} \cdots \sum_{l_{n_S-2}^{(v)}=0}^{l_{n_S-2}^{(v)}+\mu_{n_S-1}^{(v)}-1} \frac{\Gamma(l_{n_S-2}^{(v)}+\mu_{n_S-1}^{(v)})}{l_{n_S-1}^{(v)}!} \frac{\left(s+a_{n_S-1}^{(v)}\right)^{-u_{n_S-1}^{(v)}}}{(n_S/(n_S-1))^{-u_{n_S-1}^{(v)}}}$$

$$\times \Gamma\left(l_{n_S-1}^{(v)} + \mu_{n_S}^{(v)}\right)\left(s + a_{n_S}^{(v)}\right)^{-u_{n_S}^{(v)}}, \tag{13}$$

which can be re-written in a more compact form as

$$H^{(v)}(s) = \frac{c_0(m/\bar{\gamma})^{mgn_S}}{\Gamma(mg)^{n_S}} \sum_{\boldsymbol{P}^{(v)}} \sum_{\boldsymbol{T}^{(v)}} \sum_{\boldsymbol{R}^{(v)}} \sum_{\boldsymbol{L}^{(v)}} \left\{\prod_{k=1}^{n_S} c_{p,k}^{(v)} c_{tr,k}^{(v)} c_{l,k}^{(v)}\right\} \left\{\prod_{k=1}^{n_S}\left(s+a_k^{(v)}\right)^{-u_k^{(v)}}\right\}, \tag{14}$$

where $a_k^{(v)} = \frac{mn_S}{k\bar{\gamma}} \sum_{j=1}^{k}(1+t_j^{(v)})$ and the parameters $c_{l,k}^{(v)}$ and $u_k^{(v)}$ are defined as:

$$c_{l,k}^{(v)} = \begin{cases} \frac{\Gamma(\mu_1^{(v)})}{l_1^{(v)}!}(n_S)^{u_1^{(v)}} & k = 1, \\ \frac{\Gamma(l_{k-1}^{(v)}+\mu_k^{(v)})}{l_k^{(v)}!}\left(\frac{n_S}{k}\right)^{u_k^{(v)}} & k = 2,3,\ldots,n_S-1, \\ \Gamma(l_{n_S-1}^{(v)}+\mu_{n_S}^{(v)}) & k = n_S, \end{cases} \qquad u_k^{(v)} = \begin{cases} \mu_1^{(v)} - l_1^{(v)} & k = 1, \\ \mu_k^{(v)} - l_1^{(v)} + l_{k-1}^{(v)} & k = 2,3,\ldots,n_S-1, \\ \mu_{n_S}^{(v)} - l_{n_S-1}^{(v)} & k = n_S. \end{cases}$$

In (14), the summation defined with the index $\boldsymbol{L}^{(v)}$ denotes $(n_S - 1)$-fold summations with the indices $l_k^{(v)}$. Afterwards, performing the inverse Laplace transform over (14) will result in the univariate representation of the joint PDF in (11) as

$$f^{(v)}(x) = \mathcal{K}_0^{(v)} \sum_{\boldsymbol{P}^{(v)}} \sum_{\boldsymbol{T}^{(v)}} \sum_{\boldsymbol{R}^{(v)}} \sum_{\boldsymbol{L}^{(v)}} \mathcal{K}_1^{(v)} \mathcal{L}^{-1}\left\{\prod_{k=1}^{n_S}\left(s+a_k^{(v)}\right)^{-u_k^{(v)}}\right\}, \tag{15}$$

where $\mathcal{K}_0^{(v)} = \frac{c_0(m/\bar{\gamma})^{mgn_S}}{\Gamma(mg)^{n_S}}$, $\mathcal{K}_1^{(v)} = \left\{\prod_{k=1}^{n_S} c_{p,k}^{(v)} c_{tr,k}^{(v)} c_{l,k}^{(v)}\right\}$. The inverse Laplace transform can be obtained by using [28, (2.1.4-8)] that is valid for the condition $a_i^{(v)} \neq a_j^{(v)}$, $\forall i \neq j$. In order to utilize this identity and obtain the PDF expression, we rename the coefficients $a_k^{(v)}$, $k = 1,2,\ldots,n_S$, as $\tilde{a}_d^{(v)}$, $d = 1,2,\ldots,D^{(v)} \leq n_S$, such that all the coefficients $\left(\tilde{a}_d^{(v)}\right)$ are different from each other. Hence, the coefficients with equal values will result in the coefficient with the new power $\left(\tilde{u}_d^{(v)}\right)$ that is the sum of the powers of the equal-valued coefficients (i.e., $\tilde{u}_1^{(v)} = u_1^{(v)} + u_2^{(v)}$ if $a_1^{(v)} = a_2^{(v)}$). Thus, by considering this rearrangement, the PDF in (15) can be obtained as

$$f^{(v)}(x) = \mathcal{K}_0^{(v)} \sum_{P^{(v)}} \sum_{T^{(v)}} \sum_{R^{(v)}} \sum_{L^{(v)}} \mathcal{K}_1^{(v)} \sum_{k^{(v)}=1}^{D^{(v)}} \sum_{q^{(v)}=1}^{\tilde{u}_k^{(v)}} \frac{A_{kq}(-\tilde{a}_k^{(v)})}{\left(\tilde{u}_k^{(v)} - q^{(v)}\right)! (q^{(v)} - 1)!} x^{\tilde{u}_k^{(v)} - q^{(v)}} e^{-x\tilde{a}_k^{(v)}}. \quad (16)$$

$A_{kq}(-\tilde{a}_k^{(v)})$, the residue coefficients in (16), have been defined in [28, (2.1.4-8)] in detail. Integrating the PDF in (16), we can derive the CDF of the $v^{\text{th}}$ diversity branch as

$$F^{(v)}(x) = \mathcal{K}_0^{(v)} \sum_{P^{(v)}} \sum_{T^{(v)}} \sum_{R^{(v)}} \sum_{L^{(v)}} \mathcal{K}_1^{(v)} \sum_{k^{(v)}=1}^{D^{(v)}} \sum_{q^{(v)}=1}^{\tilde{u}_k^{(v)}} \mathcal{K}_2^{(v)} \psi(\tilde{u}_k^{(v)} - q^{(v)} + 1, x\tilde{a}_k^{(v)}), \quad (17)$$

where $\mathcal{K}_2^{(v)} = \frac{A_{kq}(-\tilde{a}_k^{(v)})(\tilde{a}_k^{(v)})^{q^{(v)} - \tilde{u}_k^{(v)} - 1}}{\left(\tilde{u}_k^{(v)} - q^{(v)}\right)! (q^{(v)} - 1)!}$. Also, by using the relation given in [25, (8.351.2)] as $\psi(c_1, c_2 t) = \frac{1}{c_1} e^{-c_2 t} (c_2 t)^{c_1} {}_1F_1(1; 1 + c_1; c_2 t)$, the CDF in (17) can be rewritten in terms of the confluent hypergeometric function ${}_1F_1$ [25, (210.1)] as

$$F^{(v)}(x) = \mathcal{K}_0^{(v)} \sum_{P^{(v)}} \sum_{T^{(v)}} \sum_{R^{(v)}} \sum_{L^{(v)}} \mathcal{K}_1^{(v)} \sum_{k^{(v)}=1}^{D^{(v)}} \sum_{q^{(v)}=1}^{\tilde{u}_k^{(v)}} \mathcal{K}_3^{(v)}$$

$$\times e^{-x\tilde{a}_k^{(v)}} x^{\tilde{u}_k^{(v)} - q^{(v)} + 1} {}_1F_1\left(1; \tilde{u}_k^{(v)} - q^{(v)} + 2; x\tilde{a}_k^{(v)}\right), \quad (18)$$

where $\mathcal{K}_3^{(v)} = \frac{A_{kq}(-\tilde{a}_k^{(v)})}{\left(\tilde{u}_k^{(v)} - q^{(v)} + 1\right)! (q^{(v)} - 1)!}$.

After the derivation of the statistical properties of branch SNRs as shown above, the distribution of the output SNR can be easily obtained. By using (7) and the highest order statistics [26], the CDF of the output SNR can be easily obtained by the product of the identical marginal CDF expressions $F^{(v)}(x)$ given in (17) related to the instantaneous branch SNRs $\gamma^{(v)}$:

$$F_{\gamma_U}(x) = \mathcal{K}_0 \sum_{P}\sum_{T}\sum_{R}\sum_{L} \mathcal{K}_1 \sum_{K}\sum_{Q} \mathcal{K}_2 \left\{ \prod_{v=1}^{N} \psi\left(\tilde{u}_k^{(v)} - q^{(v)} + 1, x\tilde{a}_k^{(v)}\right) \right\}, \qquad (19)$$

where $\mathcal{K}_0 = \left\{\prod_{v=1}^{N}\mathcal{K}_0^{(v)}\right\}$, $\mathcal{K}_1 = \left\{\prod_{v=1}^{N}\mathcal{K}_1^{(v)}\right\}$ and $\mathcal{K}_2 = \left\{\prod_{v=1}^{N}\mathcal{K}_2^{(v)}\right\}$. In (19), the summations defined with the indices by $P$, $T$, $R$, $L$, $K$ and $Q$ all denote $N$-fold summations with the indices $P^{(v)}$, $T^{(v)}$, $R^{(v)}$, $L^{(v)}$, $k^{(v)}$ and $q^{(v)}$, respectively. Similarly, by using (18), the CDF of the output SNR can also be expressed in terms of the confluent hypergeometric function as

$$F_{\gamma_U}(x) = \mathcal{K}_0 \sum_{P}\sum_{T}\sum_{R}\sum_{L} \mathcal{K}_1 \sum_{K}\sum_{Q} \mathcal{K}_3 \frac{x^{N+\sum_{v=1}^{N}\left(\tilde{u}_k^{(v)} - q^{(v)}\right)}}{e^{x\sum_{v=1}^{N}\tilde{a}_k^{(v)}}} \left\{ \prod_{v=1}^{N} {}_1F_1\left(1; \tilde{u}_k^{(v)} - q^{(v)} + 2; x\tilde{a}_k^{(v)}\right) \right\}, \qquad (20)$$

where $\mathcal{K}_3 = \left\{\prod_{v=1}^{N}\mathcal{K}_3^{(v)}\right\}$. By simply differentiating $F_{\gamma_U}(x) = \prod_{v=1}^{N} F^{(v)}(x)$ with respect to $x$, the PDF of the output SNR can also be obtained as $f_{\gamma_U}(x) = Nf^{(v)}(x)\prod_{v=1}^{N-1} F^{(v)}(x)$ in terms of the same coefficients, incomplete Gamma function and the confluent hypergeometric function.

### III. OUTAGE PROBABILITY ANALYSES

Outage probability which is defined as the probability that the instantaneous capacity is less than a given capacity (bandwidth efficiency) $\mathcal{R}$ (bit/s/Hertz) [1] is a useful metric while examining the overall system capacity and therefore, it comprises a realistic view on the system capacity. Since the instantaneous capacity is defined as $C_{\gamma_U} = \log_2(1 + \gamma_U)$, outage probability can be written as $P_{out} = \Pr\{\log_2(1 + \gamma_U) \leq \mathcal{R}\} = \Pr\{\gamma_U \leq 2^{\mathcal{R}} - 1\} = F_{\gamma_U}(2^{\mathcal{R}} - 1)$. As a result, the outage probability of the unified diversity scheme can be easily obtained for any TASCs by simply substituting $2^{\mathcal{R}} - 1$ into the CDF expressions in (19) and (20), respectively. Note that, although (1) and (3) are given for BER/SERs of the investigated schemes, replacing outage probability expressions instead of the BER/SER expressions will result in the average outage probability of the diversity scheme for any system configurations.

IV. ERROR PERFORMANCE ANALYSES

Based on the statistics of the output SNR which are examined up until this point, this section derives the unified expression that covers the MGF expression and BER/SER expressions of the investigated diversity schemes which are the most effective tools for examining the performance of communications systems. For this purpose, we focus on the unified integral representation given in [30, (26)] as

$$\mathcal{J}_{\gamma_U}(\theta, \epsilon, \varphi) = \theta \int_0^\infty x^\epsilon e^{-\varphi x} F_{\gamma_U}(x) dx. \tag{21}$$

Here, setting different values in the parameters of this function $(\theta, \epsilon, \varphi)$ will result in the MGF and the BER/SER of some binary/$M$-ary modulations as defined in [30]. In addition to this unified function, we define and use another unified function as given below in order to cover the SER expressions of some $M$-ary modulations as well:

$$\hat{\mathcal{J}}_{\gamma_U}(\theta, \varphi) = \theta \int_0^\infty e^{-\varphi x} F_{\gamma_U}(x) \, _1F_1\left(1; \frac{3}{2}; \frac{\varphi x}{2}\right) dx. \tag{22}$$

By considering both (21) and (22), it is possible to evaluate the MGF of the output SNR by using the identity $\mathcal{M}_{\gamma_U}(s) = \mathcal{J}_{\gamma_U}(s, 0, s)$, the average BER of BPSK, CBFSK, NCBFSK and DBPSK signals and the approximate SER of $M$-PSK signals by using the identity $P_{s,U} = \mathcal{J}_{\gamma_U}\left(\frac{\lambda_3(\lambda_2^{\lambda_1})}{2\Gamma(\lambda_1)}, \lambda_1 - 1, \lambda_2\right)$, the average SER of $M$-PAM signals by using the identity $P_{s,U} = \mathcal{J}_{\gamma_U}\left(\sqrt{\frac{3(M-1)}{\pi M^2(M+1)}}, -\frac{1}{2}, \frac{3}{M^2-1}\right)$ and the average SER of QPSK and $M$-QAM signals by using the identity $P_{s,U} = \mathcal{J}_{\gamma_U}\left(\sqrt{\frac{\lambda_4}{8\pi}}(\lambda_5 - \lambda_6), -\frac{1}{2}, \frac{\lambda_4}{2}\right) + \hat{\mathcal{J}}_{\gamma_U}\left(\frac{\lambda_4 \lambda_6}{2\pi}, \lambda_4\right)$. The parameters for the binary modulations and $M$-PSK signals are defined as $(\lambda_1, \lambda_2, \lambda_3) = (0.5, 1, 1)$ for BPSK, $(\lambda_1, \lambda_2, \lambda_3) = (0.5, 0.5, 1)$ for CBFSK, $(\lambda_1, \lambda_2, \lambda_3) = (1, 0.5, 1)$ for NCBFSK, $(\lambda_1, \lambda_2, \lambda_3) = (1, 1, 1)$ for

DBPSK and $(\lambda_1, \lambda_2, \lambda_3) = (0.5, \sin^2(\pi/M), 2)$ for $M$-PSK. And the parameters for QPSK and $M$-QAM signals are defined as $(\lambda_4, \lambda_5, \lambda_6) = (1,2,1)$ and $(\lambda_4, \lambda_5, \lambda_6) = \left(\frac{3}{M-1}, 4 - \frac{4}{\sqrt{M}}, \left(2 - \frac{2}{\sqrt{M}}\right)^2\right)$, respectively.

Thus, deriving the exact expressions of (21) and (22) will lead us to obtain and examine the performance metrics of the investigated diversity schemes for several modulations. By substituting the CDF expression of the output SNR given in (20) into the unified integral representations given in (21) and (22), and rearranging the resulting integrals in terms of the Lauricella function of $n$ variables described in [31, (2.4.2)] as

$$F_A^{(n)}(a; b_1, \dots, b_n; c_1, \dots, c_n; x_1, \dots, x_n) = \frac{1}{\Gamma(a)} \int_0^\infty e^{-st} t^{a-1} \left\{ \prod_{i=1}^n {}_1F_1(b_i; c_i; x_i t) \right\} dt, \qquad (23)$$

the unified functions in (21) and (22) can be respectively expressed as

$$\mathcal{J}_{\gamma_U}(\theta, \epsilon, \varphi) = \theta \mathcal{K}_0 \sum_P \sum_T \sum_R \sum_L \mathcal{K}_1 \sum_K \sum_Q \mathcal{K}_3 \frac{\Gamma\left(1 + \epsilon + N + \sum_{v=1}^N \left(\tilde{u}_k^{(v)} - q^{(v)}\right)\right)}{\left(\varphi + \sum_{v=1}^N \tilde{a}_k^{(v)}\right)^{1+\epsilon+N+\sum_{v=1}^N \left(\tilde{u}_k^{(v)} - q^{(v)}\right)}}$$

$$\times F_A^{(N)} \left(1 + \epsilon + N + \sum_{v=1}^N \left(\tilde{u}_k^{(v)} - q^{(v)}\right); 1, \dots, 1; \tilde{u}_k^{(1)} - q^{(1)} + 2, \dots, \tilde{u}_k^{(N)} - q^{(N)} + 2 \right.$$

$$\left. ; \frac{\tilde{a}_k^{(1)}}{\varphi + \sum_{v=1}^N \tilde{a}_k^{(v)}}, \dots, \frac{\tilde{a}_k^{(N)}}{\varphi + \sum_{v=1}^N \tilde{a}_k^{(v)}}\right), \qquad (24)$$

and

$$\hat{\mathcal{J}}_{\gamma_U}(\theta, \varphi) = \theta \mathcal{K}_0 \sum_P \sum_T \sum_R \sum_L \mathcal{K}_1 \sum_K \sum_Q \mathcal{K}_3 \frac{\Gamma\left(1 + N + \sum_{v=1}^N \left(\tilde{u}_k^{(v)} - q^{(v)}\right)\right)}{\left(\varphi + \sum_{v=1}^N \tilde{a}_k^{(v)}\right)^{1+N+\sum_{v=1}^N \left(\tilde{u}_k^{(v)} - q^{(v)}\right)}}$$

$$\times F_A^{(N+1)} \left(1 + N + \sum_{v=1}^N \left(\tilde{u}_k^{(v)} - q^{(v)}\right); 1, \dots, 1; \frac{3}{2}, \tilde{u}_k^{(1)} - q^{(1)} + 2, \dots, \tilde{u}_k^{(N)} - q^{(N)} + 2 \right.$$

$$\left. ; \frac{\varphi/2}{\varphi + \sum_{v=1}^N \tilde{a}_k^{(v)}}, \frac{\tilde{a}_k^{(1)}}{\varphi + \sum_{v=1}^N \tilde{a}_k^{(v)}}, \dots, \frac{\tilde{a}_k^{(N)}}{\varphi + \sum_{v=1}^N \tilde{a}_k^{(v)}}\right). \qquad (25)$$

The Lauricella function $F_A^{(n)}(a; b_1, \dots, b_n; c_1, \dots, c_n; x_1, \dots, x_n)$ used in (24) and (25) satisfies the convergence criterion $\sum_{i=1}^n |x_i| < 1$ for possible values of its arguments and are evaluated by using the proper integral definition in [31, (2.3.3)]. Also, [6] provides an efficient way for the numerical evaluation of this function by using the Gauss-Laguerre integration method.

Note that setting the parameters given in (4)-(7) as $(g, N) = (1, n_R)$ will result in the exact expressions related to joint TRAS/STBC scheme whereas setting $(g, N) = (n_R, 1)$ will result in those of TAS/STBC scheme. Setting the parameter as $N = 1$ in (24) and (25) will simplify the exact expressions of the unified functions for TAS/STBC scheme as

$$\mathcal{J}_{\gamma_{TAS/STBC}}(\theta, \epsilon, \varphi) = \theta \mathcal{K}_0 \sum_P \sum_T \sum_R \sum_L \mathcal{K}_1 \sum_k \sum_q \mathcal{K}_3 \frac{\Gamma(2 + \epsilon + \tilde{u}_k - q)}{(\varphi + \tilde{a}_k)^{2+\epsilon+\tilde{u}_k-q}}$$
$$\times \ _2F_1\left(1; 2 + \epsilon + \tilde{u}_k - q; 2 + \tilde{u}_k - q; \frac{\tilde{a}_k}{\varphi + \tilde{a}_k}\right), \qquad (26)$$

and

$$\hat{\mathcal{J}}_{\gamma_{TAS/STBC}}(\theta, \varphi) = \theta \mathcal{K}_0 \sum_P \sum_T \sum_R \sum_L \mathcal{K}_1 \sum_k \sum_q \mathcal{K}_3 \frac{\Gamma(2 + \tilde{u}_k - q)}{(\varphi + \tilde{a}_k)^{2+\tilde{u}_k-q}}$$
$$\times F_2\left(2 + \tilde{u}_k - q; 1,1; \frac{3}{2}, 2 + \tilde{u}_k - q; \frac{\varphi/2}{\varphi + \tilde{a}_k}, \frac{\tilde{a}_k}{\varphi + \tilde{a}_k}\right), \qquad (27)$$

where $_2F_1$ [25, (3.197-3)] and $F_2$ [29, (3.35.7-1)] denote the Gauss hypergeometric function and the Appell hypergeometric function (of the second kind), respectively. Both of these functions can be easily evaluated by using well known software programs such as MATHEMATICA and MAPLE.

V. ASYMPTOTIC DIVERSITY ORDER ANALYSIS

Since exact error probability expressions given in (24)-(27) do not give much information about the diversity order, we focus on the asymptotic behavior of the error performance. For this purpose, we derive approximate expressions for the PDF/CDF of the output SNR by using the high-SNR approximation technique given in [32] which will enable us to observe the diversity order of the unified diversity scheme clearly. The relation between the asymptotic error performance of a system and the behavior of PDF of the output SNR has been well investigated in [32]. If it is possible to express the PDF of the total channel power gain $H_U = \frac{\gamma_U}{\bar{\gamma}}$ as $f_{H_U}(x) = ax^t + o(x^t), a > 0$, one can express the error performance approximately as

$$P_{e,U} = \frac{2^t a\, \Gamma(t+1.5)}{\sqrt{\pi}(t+1)}(k\bar{\gamma})^{-(t+1)} + o(\bar{\gamma}^{-(t+1)}), \tag{29}$$

for any modulation technique with a conditional error probability (CEP) $P(e|h_U)$ consisting of $Q(\sqrt{h_U k\bar{\gamma}})$ where $k$ denotes the modulation-dependent parameter. As explored from (29), the term $t+1$ denotes the asymptotic diversity order (ADO) of the overall system. Thus, deriving the value of $t$ will be enough to examine the ADO. In order to obtain the asymptotic representation of the PDF/CDF of the total channel power gain ($H_U$), the PDF/CDF of each diversity branch has been derived asymptotically. Denoting the sorted versions of the channel power gains of the $v^{\text{th}}$ branch as $W_1^{(v)} \geq W_2^{(v)} \geq \cdots \geq W_{n_T}^{(v)}$ and the total channel power gain of this branch as $H_U^{(v)} = \sum_{k=1}^{n_S} w_{n_k}^{(v)}$, and using the union bounding technique, the CDF of $H_U^{(v)}$ can be upper bounded as given below:

$$F_{H_U^{(v)}}(x) = \Pr\left(H_U^{(v)} \leq x\right) = \Pr\left(\sum_{k=1}^{n_S} w_{n_k}^{(v)} \leq x\right) \leq \sum_{k=1}^{n_S} \Pr\left(w_{n_k}^{(v)} \leq x\right) = \sum_{k=1}^{n_S} F_{n_k}(x). \tag{30}$$

Here $F_{n_k}(x)$ denotes the CDF of the $n_k^{\text{th}}$ highest order statistics. Since the upper bounded expression of the PDF of the $n_k^{\text{th}}$ highest order statistics has been given in [4, (13)] as $f_{n_k}(x) \leq A_k x^{mg(n_T - n_k + 1) - 1} + o(x^{mg(n_T - n_k + 1) - 1})$ where $A_k = \binom{n_T}{n_T - n_k + 1}\frac{n_T - n_k + 1}{\Gamma(mg)\Gamma(mg+1)^{n_T - n_k}}$, it is possible to derive the upper bounded representation of $F_{n_k}(x)$ by simply integrating this PDF expression

$$F_{n_k}(x) = B_k x^{mg(n_T - n_k + 1)} + o(x^{mg(n_T - n_k + 1)}), \tag{31}$$

where $B_k = \frac{A_k}{mg(n_T - n_k + 1)}$. Thus, $F_{H_U^{(v)}}(x)$ can be written as

$$F_{H_U^{(v)}}(x) \leq F_{n_1}(x) + F_{n_2}(x) + \cdots + F_{n_{n_S}}(x) = \left\{B_1 x^{mg(n_T - n_1 + 1)} + o(x^{mg(n_T - n_1 + 1)})\right\}$$

$$+ \left\{B_2 x^{mg(n_T - n_2 + 1)} + o(x^{mg(n_T - n_2 + 1)})\right\} + \cdots + \left\{B_{n_S} x^{mg(n_T - n_{n_S} + 1)} + o\left(x^{mg(n_T - n_{n_S} + 1)}\right)\right\}$$

$$\approx B_{n_{min}} x^{mg(n_T-n_{min}+1)} + o\left(x^{mg(n_T-n_{min}+1)}\right), \tag{32}$$

where $n_{min} = \min_{1 \leq k \leq n_S}\{n_k\}$. The last line of (32) has been written by considering the fact that the whole expression is almost dominated by the polynomial component with the greatest exponent. Using the approximate CDF expression obtained in (32), the CDF of the overall channel power gain can be derived by using the product definition $F_{H_U}(x) = \left[F_{H_U^{(v)}}(x)\right]^N = B_{n_{min}}^N x^{mgN(n_T-n_{min}+1)} + o\left(x^{mgN(n_T-n_{min}+1)}\right)$. Finally, derivative of this CDF will result in the approximate PDF expression of the overall channel power gain:

$$f_{H_U}(x) = B_{n_{min}}^N mgN(n_T - n_{min} + 1)x^{mgN(n_T-n_{min}+1)-1} + o\left(x^{mgN(n_T-n_{min}+1)-1}\right). \tag{33}$$

As a result, for the unified diversity scheme the parameters are found as $a = B_{n_{min}}^N mgN(n_T - n_{min} + 1)$ and $t = mgN(n_T - n_{min} + 1) - 1$. Thus, by substituting these parameters into (29), the asymptotic BER/SER of both schemes can be easily obtained. Since, the product $gN = n_R$ for both joint TRAS/STBC and TAS/STBC schemes, ADO of the unified scheme is derived as $mn_R(n_T - n_{min} + 1)$. Considering that $n_{min}$ has integer values satisfying $1 \leq n_{min} \leq n_T - n_S + 1$, the value of ADO of the unified scheme changes in the interval $[mn_R n_S, mn_R n_T]$, where $mn_R n_T$ is the maximum achievable ADO value of any diversity system employing $n_T$ and $n_R$ antennas in the transmit and receive ends, respectively and operating in a Nakagami-$m$ fading environment with the fading parameter $m$. The value of the ADO has been lower bounded by the product of the fading parameter $m$, the number of receive antennas $n_R$ and the number of selected antennas ($n_S$) that are used for STBC transmission.

## VI. NUMERICAL RESULTS

In order to examine the performance of both joint TRAS/STBC and TAS/STBC schemes for the erroneous feedback conditions in Nakagami-$m$ fading channels, we present some numerical results in this section consisting of theoretical and simulation results of miscellaneous system and channel conditions for unit channel power ($\Omega = 1$), total transmit energy of $E_s$ (or $E_b$, which is average energy per bit) and the average BERs specified for the feedback link: $p_e \in \{0.0001, 0.005, 0.01, 0.05, 0.1, 0.2, 0.5\}$. In Figs. 2-7, some performance metrics such as outage probability and BER/SER are depicted versus average SNR.

Since outage probability provides a realistic view on the system capacity, we present theoretical results using (19) and simulation results of the outage probability of three-branch ($n_S = 3$) joint TRAS/$G_3$-STBC scheme in Fig. 2 for $n_R = 1$ and $n_R = 2$ in Nakagami-$m$ fading environment ($m = 2$). For the average BER values of $p_e = 0.05$ and $p_e = 0.20$ and the bandwidth efficiency $\mathcal{R} = 2$ bit/s/Hertz, the effect of the change in $p_e$ on the outage probability performance (which is relevant to the system capacity performance) is depicted. It can be easily seen from the outage probability curves that the performance degradation due to increasing $p_e$ is limited by the pure STBC (i.e., $G_3$-STBC with no antenna selection) case. Note that theoretical results are in perfect agreement with simulation results.

Figs. 3-7 show the average error performances of different modulation signals for different channel and system configurations. In Fig. 3, the SER performances of TAS/$G_2$-STBC scheme using QPSK signals are given for $n_R = 3$ and several $n_T$, in Rayleigh fading channels ($m = 1$) in the presence of FEs. The SER performances of TAS/$G_3$-STBC scheme using 16-QAM signals in the presence of FEs are given in Fig. 4 for $n_R = 2$ and several $n_T$, in One-sided Gaussian fading channels ($m = 0.5$). By examining the SNR difference of the ideal performance curves and the erroneous ones, the effect of FEs on the SER performance of TAS/STBC scheme can be observed. As seen in Figs. 3 and 4, for a moderate BER of $p_e = 0.01$ for the feedback link, the SNR degradation for TAS/$G_2$-STBC $\{n_T = 3, n_R = 3\}$, TAS/$G_2$-STBC $\{n_T = 4, n_R = 3\}$, TAS/$G_3$-STBC $\{n_T = 4, n_R = 2\}$ and TAS/$G_3$-STBC $\{n_T = 5, n_R = 2\}$ is 0.14 dB, 0.36 dB, 0.45 dB and 0.8 dB, respectively at a SER of $10^{-5}$ whereas the degradation values for severe feedback channel conditions with BERs of $p_e = 0.2$ and $p_e = 0.5$ are 1.4 dB, 2.2 dB, 3.25 dB, 4.3 and 2.4 dB, 3.5 dB, 3.9 dB, 6.1 dB respectively for the same schemes. The SER performances related to the erroneous

feedback cases are upper bounded by the pure STBC (i.e., STBC with no transmit antenna selection) case for the BSC BER values satisfying $p_e \leq 0.5$. Thus, the ADO of TAS/STBC schemes in the presence of FEs would be at least equal to that of the pure STBC case (which is $mn_R n_S$).

The SER curves related to joint TRAS/$G_2$-STBC scheme using QPSK signals in the presence of FEs are given in Fig. 5 for $n_R = 3$ and several $n_T$, in Rayleigh fading channels ($m = 1$). The BER performances of joint TRAS/$G_3$-STBC scheme using CBFSK signals in the presence of FEs are given in Fig. 6 for $n_R = 2$ and several $n_T$, in Nakagami-$m$ fading channels ($m = 2$). The variation of the error performance of joint TRAS/$G_2$-STBC $\{n_T = 3\}$ scheme due to the variation of the number of receive antennas ($n_R \in \{1, 2, 3\}$) and BSC BER values ($p_e \in \{0.01, 0.1\}$) has been depicted in Fig. 7 for BPSK modulation signals and Rayleigh fading case ($m = 1$). The effect of FEs on the SER performance of joint TRAS/STBC schemes can be observed by interpreting the SNR difference of the ideal performance curves and the erroneous ones. In Figs. 5 and 6, for a moderate BER of $p_e = 0.01$ for the feedback link, the SNR degradation values for joint TRAS/$G_2$-STBC $\{n_T = 3, n_R = 3\}$, joint TRAS/$G_2$-STBC $\{n_T = 4, n_R = 3\}$, joint TRAS/$G_3$-STBC $\{n_T = 4, n_R = 2\}$ and joint TRAS/$G_3$-STBC $\{n_T = 5, n_R = 2\}$ are 0.4 dB, 1.3 dB, 0.3 dB and 0.25 dB, respectively at a SER of $10^{-5}$ whereas the degradation values for severe feedback channel conditions with BERs of $p_e = 0.2$ and $p_e = 0.5$ are 2.75 dB, 4.5 dB, 1.1 dB, 1.75 dB and 4 dB, 6 dB, 1.65 dB, 2.6 dB respectively for the same schemes. Also, the BER variation depicted in Fig. 7 shows that the SNR degradation caused by two different FE cases ($p_e = 0.01$ and $p_e = 0.1$) are decreasing for the increasing values of $n_R$. For a BER of $10^{-5}$ and BSC BER of $p_e = 0.01$, the SNR degradation values for $n_R = 1, 2$ and 3 are seen to be 1.6 dB, 1.2 dB and 1.1 dB, whereas for a BSC BER of $p_e = 0.1$, these values are 5.7 dB, 3.7 dB and 3.1 dB. The ADO of joint TRAS/STBC scheme in the presence of FEs would be lower bounded by the ADO provided by the STBC&RAS scheme (i.e., joint TRAS/STBC with no transmit antenna selection) similar to the TAS/STBC case. However, the error performance curves related to joint TRAS/STBC scheme in the erroneous feedback cases are not upper bounded by the STBC&RAS case (that can easily be seen from the scenarios in Figs. 5 and 6) for the BSC BER values for which the erroneous feedback cases are still superior to pure STBC scheme (i.e., TAS/STBC with no transmit antenna selection). The robustness of TAS/STBC scheme when compared to joint TRAS/STBC scheme against FEs can also be observed by comparing the SNR degradation values of both schemes for the same BSC BER values given in Figs 3 and 5. Also, as seen in the BER/SER curves of Figs. 3-6, increasing the total number of antennas in the transmitter of the joint TRAS/STBC scheme does not enhance the error performance for all SNR values in the presence of FEs whereas the same increment would still result in SNR gain for TAS/STBC scheme. This result reveals the sensitive selection structure that joint TRAS scheme has against the MRC-like combining structure of TAS/STBC scheme which can also be interpreted as the superiority of conventional TAS/STBC scheme in the presence of FEs when compared to the joint TRAS/STBC scheme. The differing behaviors of these schemes in the presence of FEs are heavily related to the selection criteria employed. The selection structure employed in TAS/STBC scheme causes impairments in transmit diversity in the presence of FEs while *perfectly* maintaining the receiver diversity (provided by MRC-like structure). Whereas, in the presence of FEs, the *joint* selection structure employed in joint TRAS/STBC scheme would cause impairments both at transmit and receive diversity that may result in the exclusion of the best transmit-receive links. Besides, Monte Carlo simulation results perfectly match with the theoretical results in all performance figures.

## VII. Conclusions

This paper has focused on the performance evaluation of the unified diversity scheme covering both joint TRAS/STBC and TAS/STBC schemes considering a generic fading model like Nakagami-*m* in the presence of practical impairments in the feedback link (i.e., FEs). Extensive analyses have been carried out in order to derive the exact expressions related to the statistics of the output SNR (PDF, CDF and MGF) and the capacity and error performances (outage probability and BER/SER) for both schemes in ideal and erroneous conditions. Also, by examining the asymptotic diversity order analysis of the unified scheme, we have shown that both schemes achieve full diversity order (i.e., $mn_R n_T$) for ideal conditions, while in the presence of FEs they keep maintaining at least the asymptotic diversity order provided by the STBC (with no transmit antenna selection) scheme itself and the receive diversity order provided by RAS or MRC-like structure (i.e., $mn_R n_S$). As indicated in Figs. 2-7, for ideal conditions of both schemes and the erroneous conditions of TAS/STBC scheme, the outage probability and BER/SER performances can easily achieve great improvements due to higher diversity orders yielded by the increase in the number of transmit antennas used for TAS in both schemes and/or the number of receive antennas available for RAS and MRC-like structure in joint TRAS/STBC and TAS/STBC schemes, respectively. Whereas, for erroneous cases of joint TRAS/STBC scheme, the increment in the total number of transmit antennas causes deterioration on the performance for high SNR values as seen in Figs 5 and 6. Moreover, Monte Carlo simulation results have been obtained as in perfect agreement with the theoretical results. The performance results clearly point out the superiority of TAS/STBC scheme to joint TRAS/STBC scheme especially in the presence of FEs. This superiority is provided by the virtue of multiple RF chains employed in the receiver side of TAS/STBC. However the simple receiver structure of joint TRAS/STBC (employing only a single RF chain) tends this scheme to be a more feasible technique for multi-antenna communications systems rather than TAS/STBC scheme. Also, for reasonable average BER values (such as $p_e = 0.01$) specified for the feedback link, both schemes still achieve considerable outage and error probability performances (with tolerable SNR degradation) that provide both schemes to maintain significant importance in real-world wireless communications system design.

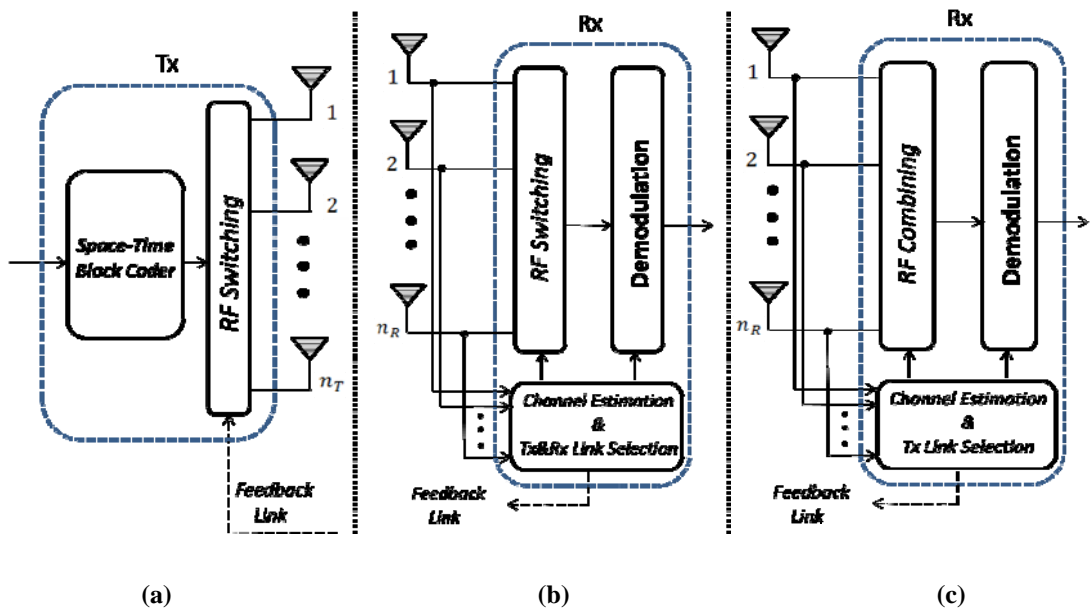

**Fig. 1** Block diagram of both schemes: (a) Transmitter of both schemes, (b) Receiver of Joint TRAS/STBC scheme, (c) Receiver of TAS/STBC scheme

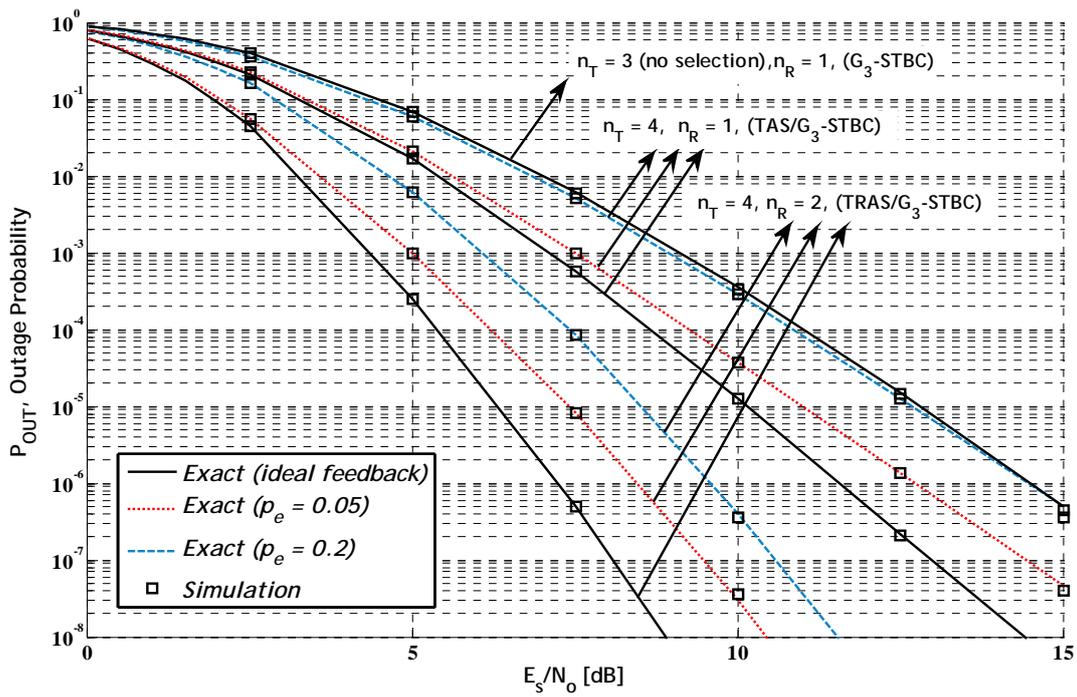

**Fig. 2** Outage probability vs. average SNR per symbol for joint TRAS/$G_3$-STBC scheme for several $n_T$, $n_R$ ($\mathcal{R} = 2$ bit/s/Hertz, $m = 2$, $p_e = 0.05$ and $p_e = 0.2$)

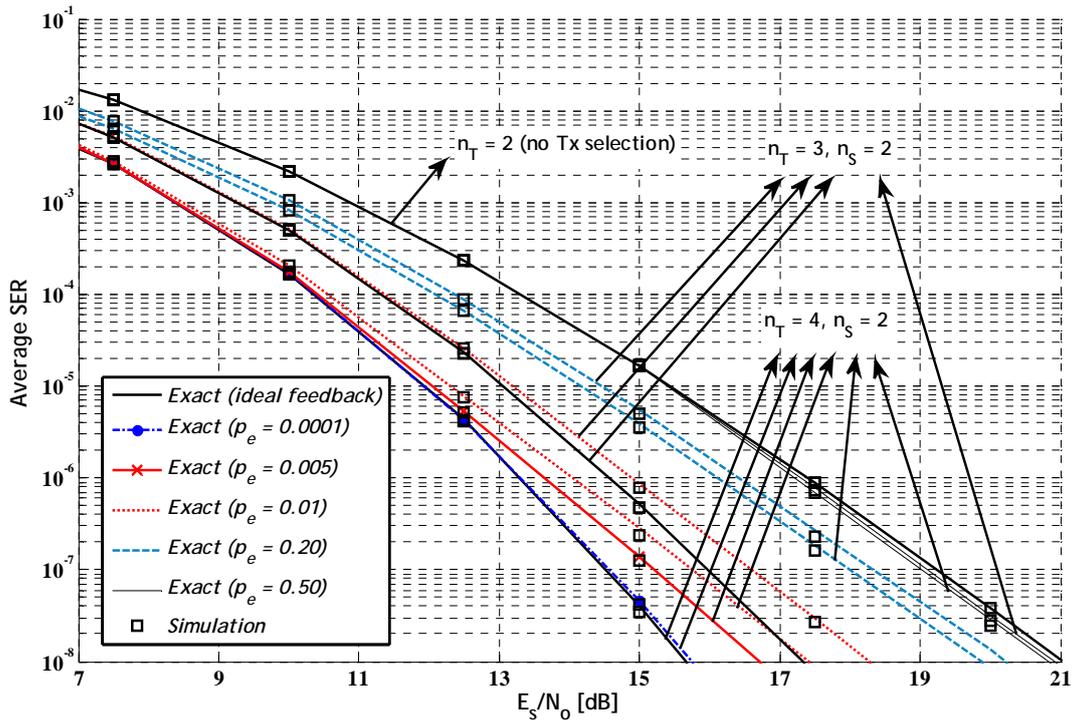

**Fig. 3** Average SER vs. average SNR per symbol for TAS/$G_2$-STBC scheme using QPSK signals for several $n_T$ and $p_e$ ($n_R = 3, m = 1$ (Rayleigh fading channels))

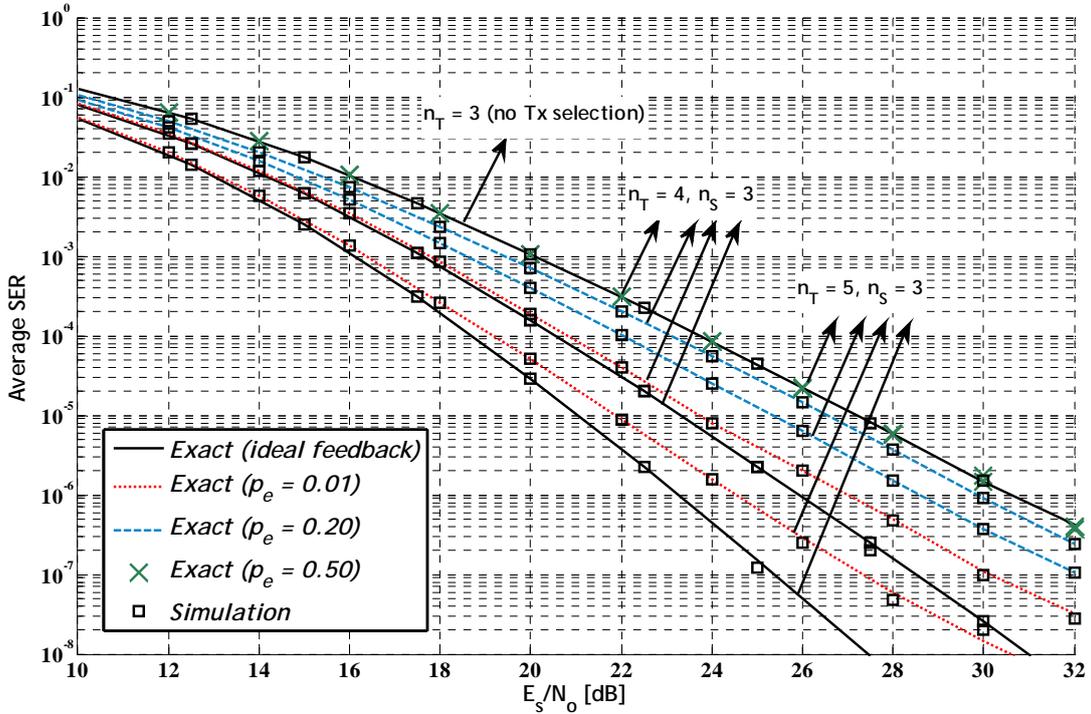

**Fig. 4** Average SER vs. average SNR per symbol for TAS/$G_3$-STBC scheme using 16-QAM signals for several $n_T$ and $p_e$ ($n_R = 2, m = 0.5$ (One-sided Gaussian fading channels))

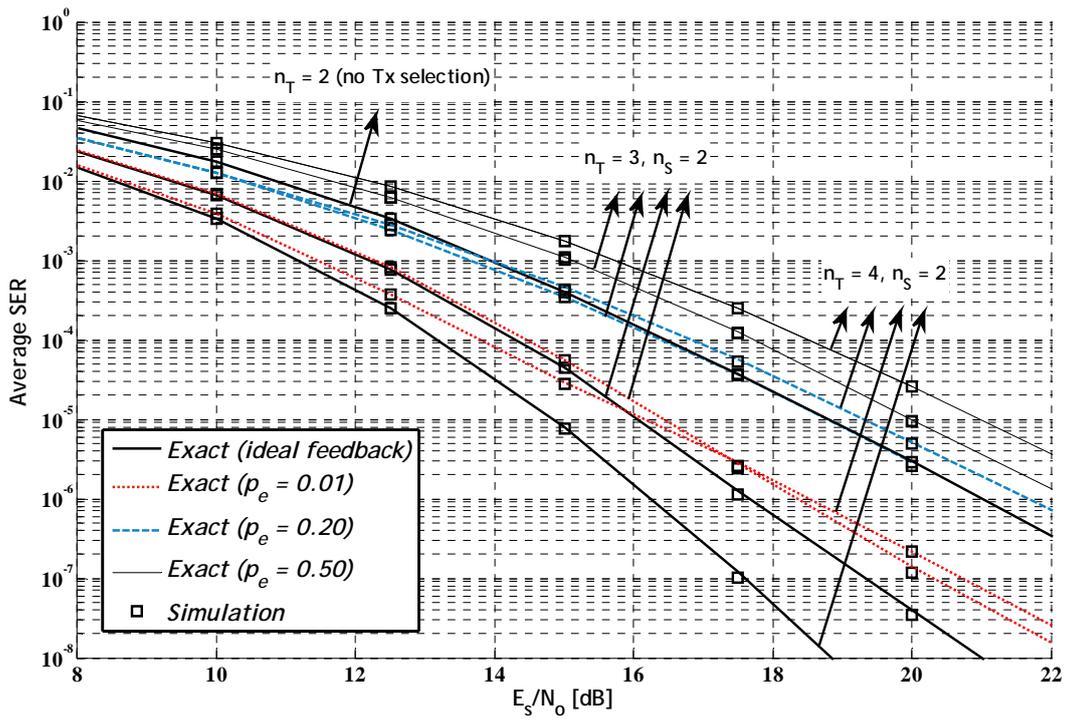

**Fig. 5** Average SER vs. average SNR per symbol for joint TRAS/G$_2$-STBC scheme using QPSK signals for several $n_T$ and $p_e$ ($n_R = 3, m = 1$ (Rayleigh fading channels))

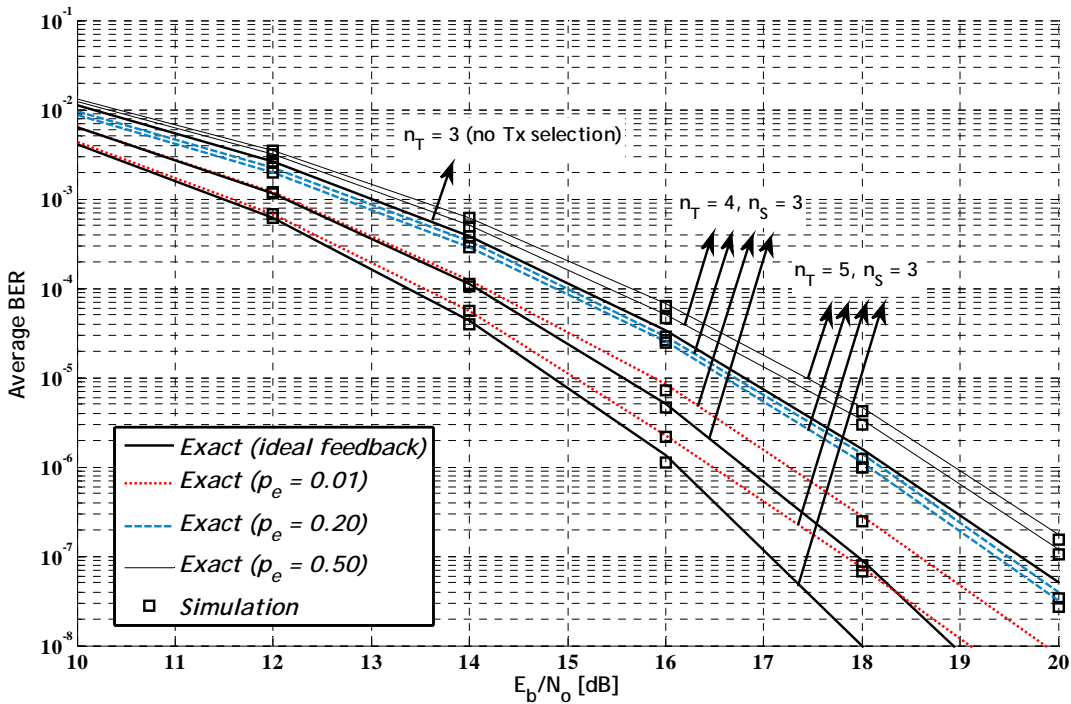

**Fig. 6** Average BER vs. average SNR per bit for joint TRAS/G$_3$-STBC scheme using CBFSK signals for several $n_T$ and $p_e$ ($n_R = 2, m = 2$)

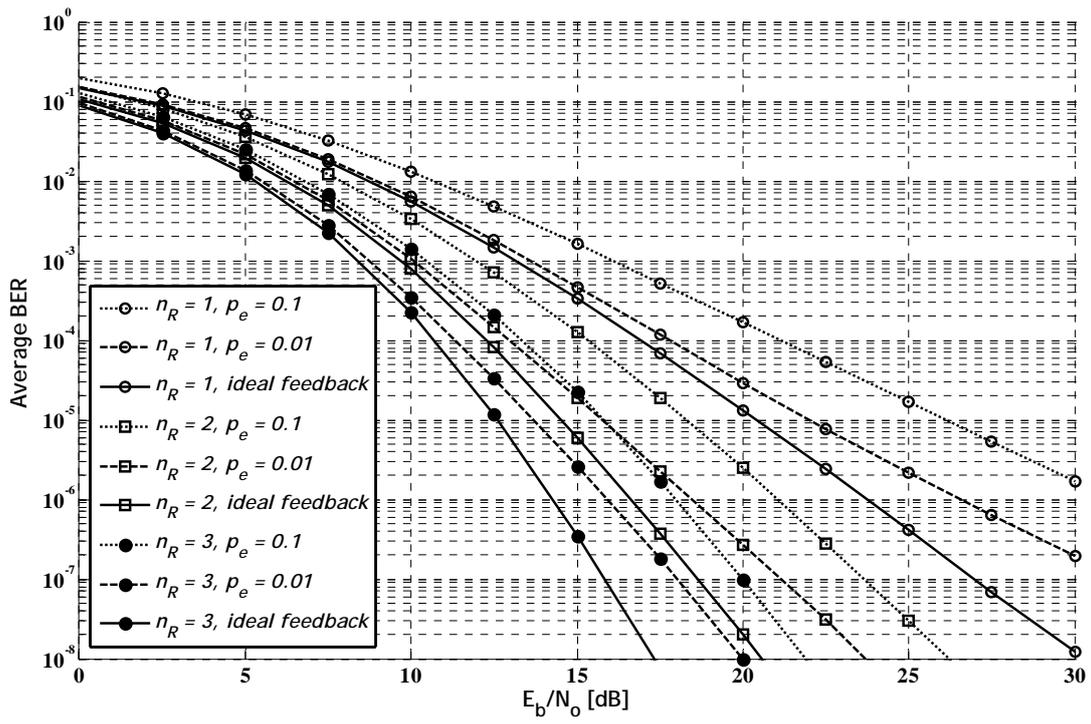

**Fig. 7** Average BER vs. average SNR per bit for joint TRAS/G$_2$-STBC ($n_T = 3, n_S = 2$) scheme using BPSK signals and Rayleigh fading channels ($m = 1$) ($p_e \in \{0.01, 0.1\}$, $n_R \in \{1, 2, 3\}$)